\documentclass[default,iicol]{sn-jnl}

\RequirePackage{filecontents}
\usepackage[T1]{fontenc}
\usepackage[latin9]{inputenc}
\usepackage{geometry}
\usepackage{float}
\usepackage{amsmath}
\usepackage{amsthm}
\usepackage{graphicx}
\usepackage{amssymb}
\usepackage{multirow}
\usepackage{babel}
\usepackage{etoolbox}
\usepackage{physics}
\usepackage{natbib}
\setcitestyle{authoryear,round,semicolon,aysep={},ayysep={,}} 

\newcommand{\qpufid}{\mathrm{qPUF}_{\mathrm{id}}}
\newcommand{\qgen}{\operatorname{QGen}(\lambda)}
\newcommand{\qeval}[1]{\operatorname{QEval}\left( #1 \right)}
\newcommand{\rout}{r_{\mathrm{out}}}
\newcommand{\rhat}{\hat{r}_{\mathrm{out}}}
\newcommand{\Uin}{U_{\mathrm{in}}}
\newcommand{\Uint}{U_{\mathrm{in}}(\Vec{\theta})}

\jyear{2021}%

\theoremstyle{thmstyleone}%
\theoremstyle{thmstyletwo}%
\theoremstyle{thmstylethree}%
\DeclareMathOperator*{\E}{\mathbb{E}}

\def\tablename{Table}

\begin{document}

\title[Article Title]{Learning Classical Readout Quantum PUFs based on single-qubit gates}


\author*[1]{\fnm{Niklas} \sur{Pirnay}}\email{n.pirnay@tu-berlin.de}

\author[1,2]{\fnm{Anna} \sur{Pappa}}\email{anna.pappa@tu-berlin.de}

\author[1,3]{\fnm{Jean-Pierre} \sur{Seifert}}\email{jpseifert@sect.tu-berlin.de}

\affil*[1]{\orgdiv{Electrical Engineering and Computer Science Department}, \orgname{Technische Universit{\"a}t Berlin}, \orgaddress{\city{Berlin}, \postcode{10587}, \country{Germany}}}

\affil[2]{\orgdiv{Fraunhofer Institute for Open Communication Systems}, \orgaddress{\city{Berlin}, \postcode{10587}, \country{Germany}}}

\affil[3]{\orgdiv{Fraunhofer Institute for Secure Information Technology}, \orgaddress{\city{Darmstadt}, \postcode{64295}, \country{Germany}}}


\abstract{Physical Unclonable Functions (PUFs) have been proposed as a way to identify and authenticate electronic devices. Recently, several ideas have been presented that aim to achieve the same for quantum devices. Some of these constructions apply single-qubit gates in order to provide a secure fingerprint of the quantum device. In this work, we formalize the class of \textit{Classical Readout Quantum PUFs} (CR-QPUFs) using the \textit{statistical query} (SQ) model and explicitly show insufficient security for CR-QPUFs based on single qubit rotation gates, when the adversary has SQ access to the CR-QPUF. We demonstrate how a malicious party can learn the CR-QPUF characteristics and forge the signature of a quantum device through a modelling attack using a simple regression of low-degree polynomials. The proposed modelling attack was successfully implemented in a real-world scenario on real IBM Q quantum machines.
We thoroughly discuss the prospects and problems of CR-QPUFs where quantum device imperfections are used as a secure fingerprint.}


\maketitle
 
\section{Introduction}

Secure attestation of cloud computing resources has been in the focus of research to create trust in the cloud, since through it, cloud computing customers can make sure that they are provided access to the correct hardware platform \citep{attestation}.
Trusted hardware is especially important for quantum computing (QC), since the high sensitivity to noise in low-grade machines can have detrimental effects on vital calculations \citep{regev_grover_noise, Preskill2018_nisq}.\\
In the current quantum computing ecosystem, quantum computers of competing hardware manufacturers are aggregated by third party cloud providers, giving customers access to a multitude of QC hardware platforms.
However, a malicious party acting as a legitimate quantum cloud provider, might try to reroute customer circuits to cheaper lower-grade hardware, to either save money, censor certain customers or damage the reputation of selected QC platforms.
It is therefore an important endeavour to ensure that quantum cloud customers can authenticate the accessed computing devices; from the customers' side to lower the risk of making extensive business decisions based on potentially corrupted results, and from the platform manufacturers' side, to safeguard against third parties trying to damage their reputation.

In the classical world, Physical Unclonable Functions (PUFs) have been proposed as a way to identify and authenticate electronic devices \citep{pappu_physical_2002,Brzuska2011,DBLP:books/sp/Maes13}. 
Recently, Quantum PUFs (QPUFs) have emerged that aim to achieve the same goals for quantum devices. The primer of \cite{vskoric2012quantum} introduced the concept of Quantum Readout PUFs (QR-PUFs), which were later generalized and formalized in the framework of \cite{doosti2020client} and rigorously analyzed by \cite{arapinis2021quantum}.
However, the authentication protocols using QPUFs require a quantum memory and a quantum communication channel between the verifier and prover to exchange quantum states.
Other studies propose to use QPUFs as generators for cryptographic keys to establish information theoretically safe communication \citep{horstmeyer_physical_2013, nikolopoulos_remote_2021}, however they are built with special-purpose hardware, that does not fit the gate-based computing model of QC cloud platforms.
The recent work by \cite{phalak_quantum_2021} proposes a QPUF, where the authentication protocol requires only classical communication, no quantum memory and can be implemented on gate-based quantum computers to fingerprint these devices.

In this work, we adapt the QPUF framework of \cite{doosti2020client} in order to address the case of Classical Readout Quantum PUFs (CR-QPUFs) where verifier and prover communicate classically to authenticate a quantum device.
CR-QPUF constructions in essence rely on classically communicating a unitary transformation, the challenge, which is performed on the quantum device. The device-specific imperfections degenerate the unitary challenge and result in a unique output state, which is measured to obtain the response. 
Interestingly, due to the prevalent noise in Noisy Intermediate-Scale Quantum (NISQ) devices, we find that in order to build a robust CR-QPUF, responses are designed as statistical queries to the QPUF. In fact, the so-called Hadamard CR-QPUF \citep{phalak_quantum_2021} intrinsically uses the statistical query model, which we will present in this paper.
Even though it has been shown that restrictions to the SQ model can be used to obtain unlearnability results \citep{learning-local-quant-dist, gollakota2021hardness, kearns98}, we present a successful learning attack on the Hadamard CR-QPUF in the SQ model. We thereby show that the remote authentication scheme using the Hadamard CR-QPUF is not secure against learning attacks. We show that an attacker is able to model and predict the Hadamard CR-QPUF characteristics and hence forge the quantum device fingerprint using machine learning. Additionally, we investigate natural extensions of the Hadamard CR-QPUF and observe similar security flaws.

This work is concluded with an in-depth discussion of the prospects and drawbacks of CR-QPUFs. It is possible that in the NISQ era, constructing CR-QPUFs in the Statistical Query (SQ) model might provide security guarantees against learning attacks. However, there are many open questions that need to be addressed. How can imperfections be modelled? Are entangled resources necessary? What security guarantees can we get from the SQ model? We discuss these questions based on our findings regarding the insecurity of Hadamard CR-QPUFs. Additionally, we discuss the security of CR-QPUFs in the contrarian views of the power of NISQ devices, i.e. whether output distributions of NISQ devices can in general be modelled using low-degree polynomials or not \citep{kalai2020argument}.
Finally, we propose future work and possible next steps for investigating CR-QPUFs that are secure against machine learning attacks.

\section{Classical-Readout QPUFs}
\label{sec:cr-qpuf}
In general, PUFs exhibit unique input-output relations which depend on the physical properties of the device on which the PUF is implemented.
These unique relations are commonly referred to as \textit{challenge-response pairs} (CRPs) and constitute a fingerprint of the physical device.

\begin{figure}
    \centering
    \includegraphics[scale=.4]{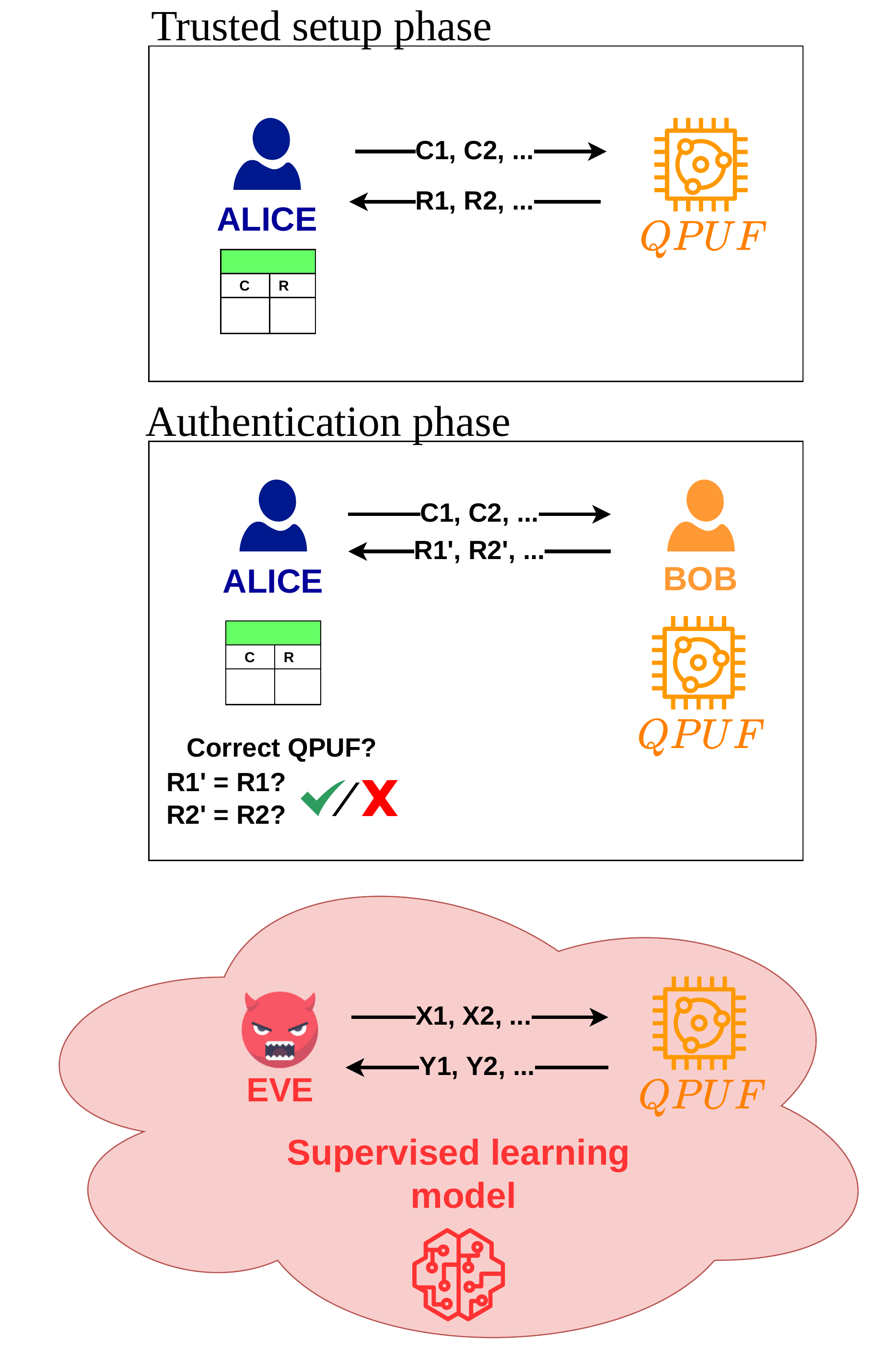}
    \caption{Schematic illustration of the authentication protocol and attacker model considered. In the trusted setup phase, Alice will create a CRP database by repeatedly invoking the QPUF with random challenges. This database will serve as a unique fingerprint of the QPUF. In the authentication phase, Bob (who claims to be in possession of the QPUF) is required to answer with the correct responses to challenges in Alice's CRP database. In the modelling attack, Eve has access to the QPUF and aims to learn a model of the input-output behaviour in order to predict responses to future challenges, thus succeeding in the authentication phase using solely the model of the QPUF.}
    \label{fig:puf_protocol}
\end{figure}

Authentication schemes based on PUFs typically use a challenge and response protocol between a verifier and a prover. The goal of the prover (here called Bob) is to prove to the verifier (here called Alice) that he is in possession of the PUF, by demonstrating the ability to query the PUF.
In figure \ref{fig:puf_protocol} we schematically depict the authentication protocol of a QPUF that we consider in this work.
Before the authentication can take place, a trusted setup phase is required, during which Alice can directly interact with the QPUF. This enables her to build a secret CRP database by repeatedly querying the QPUF with randomly chosen challenges and storing the challenges and respective responses in the database. At a later stage, when Bob wants to prove possession of the QPUF to Alice, he will need to provide the correct corresponding response to a CRP chosen randomly by Alice from her database.
If the response matches the CRP entry in the secret database of Alice, Alice authenticates that Bob is in possession of the QPUF. An honest Bob will query the QPUF with the challenge and send the respective response back. A dishonest Bob can try to implement an attack, for example a \textit{modelling attack}, where by figuring out the QPUF's input-output characteristics he can predict the correct response to arbitrary challenges without possessing the QPUF.  

The protocol described above can be used to authenticate QCs provided by third party cloud platforms. During the trusted setup phase, the user (Alice) obtains a (certified) CRP database of a QPUF that is implemented on the quantum device of a certain manufacturer. If Alice wants to authenticate the QC at a later point when she no longer has physical access to the hardware, she queries the QPUF again with challenges from her CRP database and compares the respective responses.

Prior work of \cite{vskoric2012quantum} introduced the concept of a Quantum-Readout PUF (QR-PUF), where challenges and responses are communicated using a quantum channel, thereby presenting a protocol to authenticate a QR-PUF without the need to rely on a trusted readout device. \cite{doosti2020client} proposed a generalised framework for QPUFs, where a secret unitary transformation is performed on challenge states. Since the challenges and responses are quantum states, they need to be stored in a quantum memory; unfortunately, this is not easy to achieve in practice.

A recent proposal by \cite{phalak_quantum_2021} introduced the concept of a Classical-Readout QPUF (CR-QPUF) where a quantum device is queried classically, removing the requirement for a quantum memory. In the proposed protocol, the challenge is a classical description of a parameterized unitary that is run on the quantum computer. The response is the mean of multiple measurement outcomes of the qubits in the computational basis. Both the challenge unitary and the mean value with finite samples, are communicated classically between verifier and prover. The goal of the protocol is to leverage imperfections in the quantum computer as hidden parameters that identify the quantum device uniquely.
Table \ref{tab:qpuf-class} summarises the different categories of QPUF protocols at the quantum-classical intersection.
\begin{table}
\caption{\label{tab:qpuf-class} Classification of (Quantum) PUFs}
\begin{center}
\bgroup
\def\arraystretch{1}
  \hskip-0.51cm
  \begin{tabular}{ccccc}
        & & \multicolumn{2}{ c }{Device}  \\ \cline{3-4}
        & & Quantum & Classical \\ \cline{2-4}
        \multicolumn{1}{ c  }{\multirow{4}{*}{CRPs} } &
        \multicolumn{1}{ c }{Quantum} & QPUF & QR-PUF \\
        \multicolumn{1}{ c }{} & {} & \citep{doosti2020client} & \citep{vskoric2012quantum} \\
        \cline{2-4}
        \multicolumn{1}{ c  }{}                        &
        \multicolumn{1}{ c }{Classical} & {CR-QPUF} & PUF \\
        \multicolumn{1}{ c }{} & {} & {} & \citep{DBLP:books/sp/Maes13} \\
        \cline{2-4}
    \end{tabular}
\egroup
\end{center}
\end{table}\\
Here, we introduce the formalism of CR-QPUFs for NISQ devices according to the QPUF framework of \cite{doosti2020client}.
NISQ devices are subject to noise which consists of systematic noise and non-systematic noise. In the following, we refer to systematic noise as device imperfections and non-systematic noise as white noise. We will generalize the concept of CR-QPUFs envisioned by \cite{phalak_quantum_2021} and accurately describe the challenge-response space to enable further rigid analysis.

We will denote with $\qpufid$ the unique identifying properties of a quantum device with identity $\mathrm{id}$, generated by the process $\qgen$, where $\lambda$ is the security parameter:
\begin{equation*}
\qpufid \leftarrow \qgen    
\end{equation*}
Let $\mathcal{U}$ be a subspace of unitary transformations, called the \textit{challenge space}. Given a challenge $\Uin \in \mathcal{U}$, the response $\rout$ is a set of i.i.d. samples from running $\Uin$ on the quantum computer identified by $\qpufid$ and measuring in the computational basis.
Thus, the completely-positive trace-preserving map $\Lambda_{\mathrm{id}}$ that is performed by the quantum device $\mathrm{id}$ with properties $\qpufid$ is a function of the challenge unitary run on the device:
\begin{equation*}
\Lambda_{\mathrm{id}}(\Uin) =: \Lambda^{\mathrm{id}}_{\mathrm{in}}    
\end{equation*}
The density matrix $\rho_{\mathrm{in}}^{\mathrm{id}}$ is then defined as the result of performing the quantum operation to the $\ket{0}$ state, i.e,
\begin{equation*}
\rho_{\mathrm{in}}^{\mathrm{id}} = \Lambda^{\mathrm{id}}_{\mathrm{in}} \ket{0}\bra{0} {\Lambda^{\mathrm{id}\dagger}_{\mathrm{in}}}\text{.}    
\end{equation*}
Note that $\rho_{\mathrm{in}}^{\mathrm{id}}$ does not equal the state after perfectly applying $\Uin$, but a mutated version thereof, where the state has been altered depending on the device imperfections.
Further, we define the \textit{Born distribution} as
\begin{equation*}
    \mathcal{P}_{\Uin,\mathrm{id}}(x) = Tr(M_{x} \rho_{\mathrm{in}}^{\mathrm{id}} M_{x}^{\dagger})    
\end{equation*}
which is the distribution over $n$-bit strings, induced by measuring the qubits in the computational basis. Here $M_{x}$ corresponds to the measurement operator of $\ket{x}$.
The generic evaluation $\mathrm{QEval}$ of the QPUF is then simply sampling from the Born distribution. That is,
\begin{equation*}
   \rout \leftarrow \qeval{\qpufid, \Uin}  
\end{equation*}
and for noise-free devices we obtain
\begin{equation*}
   \rout = \left \{  x : x \sim \mathcal{P}_{\Uin,\mathrm{id}} \right \} \text{.} 
\end{equation*}
However, in NISQ devices, white noise is prevalent and the samples in $\rout$ are corrupted by random noise. We model the random noise as random bit flips. 
Analog to noisy boolean functions \citep{benjamini1999noise}, given $x, y \in \{0,1\}^n$ and $\epsilon \in (0, 1)$, let $y$ be a random perturbation of $x$, i.e. $x_i = y_i$ with probability $1-\epsilon$, independently for distinct $i$'s. We denote the random perturbation by $\mathrm{N}_{\epsilon}(x) = y$. Thus for NISQ devices, the response $\hat{r}_\mathrm{out}$ for a given challenge $\Uin$ is the finite set of noisy samples,
\begin{equation*}
   \hat{r}_\mathrm{out} = \left \{  \mathrm{N}_{\epsilon}(x) : x \sim \mathcal{P}_{\Uin,\mathrm{id}} \right \} \text{.}   
\end{equation*}
Thus, in order to construct a robust CR-QPUF response on such noisy devices, it is intuitive to use global statistical properties across multiple samples \citep{benjamini1999noise}. The Statistical Query (SQ) model \citep{kearns98} provides an excellent formalism to describe this mechanism. In fact the Hadamard CR-QPUF, which is described in the next section, naturally uses the SQ model to build a robust QPUF authentication scheme.

In the SQ model for CR-QPUFs, given an efficiently computable function $\phi: \{ 0,1 \}^{n} \rightarrow \{ 0,1 \}^{n}$ the QPUF evaluation $\mathrm{QEval}$ responds with some $\Vec{R}_{\mathrm{out}}$ that is $\tau$-close to the expectation value of $\phi$. Thus, in the SQ model, we have that
\begin{equation*}
    \Vec{R}_{\mathrm{out}} \leftarrow \qeval{\qpufid, \Uin}    
\end{equation*}
where
\begin{equation*}
    \abs{ \E_{x \sim \mathcal{P}_{\Uin, \mathrm{id}}} [ \phi(x) ] - \Vec{R}_{\mathrm{out}} } \leq \tau \text{.}  
\end{equation*}\\
Note that behind the scenes, the noise-robust response $\Vec{R}_{\mathrm{out}}$ is constructed from noisy samples in $\hat{r}_\mathrm{out}$.  In fact, Chernoff-Hoeffding bounds imply that for noise rate $\epsilon < 1/2$, for any efficiently-computable function $\phi$, the SQ model can be simulated using noisy samples $\hat{r}_\mathrm{out}$, where $\abs{\hat{r}_\mathrm{out}}$ is polynomial in $n$, $\tau^{-2}$ and $\epsilon^{-2}$.\\
It follows that any algorithm that tries to learn the distribution $\mathcal{P}_{\Uin, \mathrm{id}}$, while being restricted to using the SQ model, is less (or equally) powerful than an algorithm that has direct access to $\hat{r}_\mathrm{out}$. In the next section, we will consolidate the Hadamard CR-QPUF in this framework.

\section{The Hadamard CR-QPUF}

\begin{figure}[]
\centering
\includegraphics[]{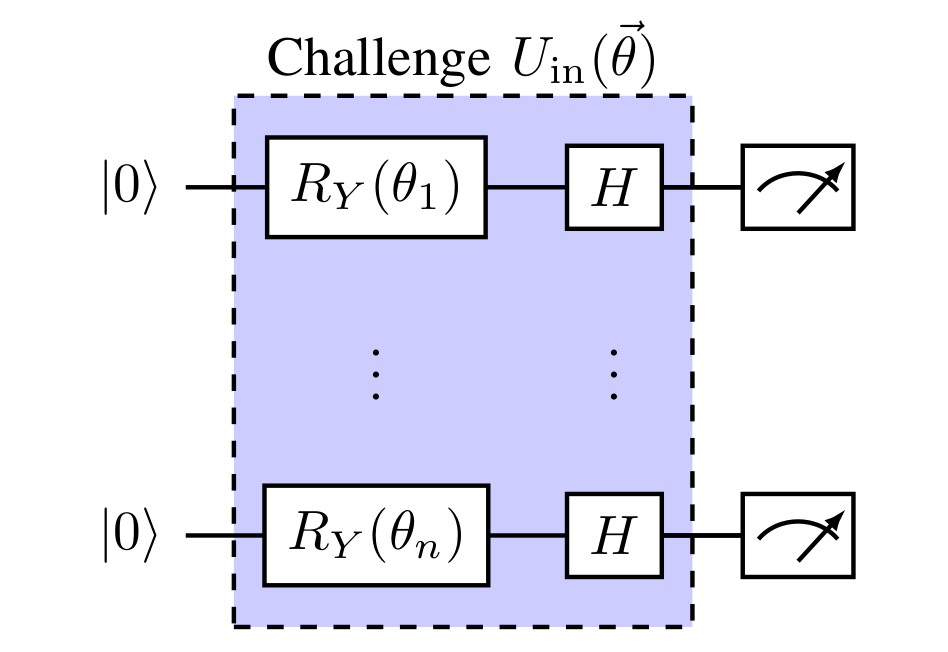}
\caption{Circuit of the parameterized challenge unitary family in the Hadamard CR-QPUF. The challenges are run on a $n$-bit quantum computer, returning multiple shots per challenge.}
\label{fig:h-qupuf_circuit}
\end{figure}

The Hadamard CR-QPUF introduced by \cite{phalak_quantum_2021} aims at robustly authenticating a quantum computer using device-specific qubit imperfections via a classical communication channel. The CR-QPUF challenges are given by a parameterized quantum circuit that includes parameterized single qubit rotations and Hadamard gates. The responses are the empirical mean of projecive meaurements of the qubits in the computational basis. In the following, we will define the Hadamard CR-QPUF in the framework described above using the SQ model.\\
The class of challenge unitaries $\mathcal{U}$ that is used in the Hadamard CR-QPUF is depicted in figure \ref{fig:h-qupuf_circuit} and given by:
\begin{equation*}
    \mathcal{U} := \left \{ \bigotimes_{i=1}^{n} H R_{Y}(\theta_i) : \theta_i \in [0, 2\pi).\forall i \in \{1,...,n\}  \right \}
\end{equation*}
Let $\qpufid$ be the properties of the quantum computer $\mathrm{id}$ and let $\Uint \in \mathcal{U}$ be a chosen challenge unitary acting on $n$ qubits that is parameterized by rotations $\Vec{\theta}$.
The response $\Vec{R}_{\mathrm{out}}$ is given by the $\tau$-close approximation to the expectation value of the qubits, more precisely:
\begin{equation*}
    \abs{ \E_{x \sim \mathcal{P}_{\Uin, \mathrm{id}}} [ x ] - \Vec{R}_{\mathrm{out}} } \leq \tau \text{ .}
\end{equation*}
Thus, the response $\Vec{R}_{\mathrm{out}}$ is the result of a statistical query as introduced in section \ref{sec:cr-qpuf}, where the function $\phi$ is the identity. Following from the law of large numbers, the response $\Vec{R}_{\mathrm{out}}$ is computed using noisy samples $\hat{r}_\mathrm{out}$ that are obtained by sampling from $\mathcal{P}_{\Uint, \mathrm{id}}$. Thus the QPUF response $\Vec{R}_{\mathrm{out}}$ is given by:
\begin{equation*}
    \Vec{R}_{\mathrm{out}} = \frac{1}{\abs{\rhat}}\sum_{r \in \rhat}  r  \text{, subject to} 
\end{equation*}
\begin{equation*}
    \epsilon \leq \tau \abs{ \E_{x \sim \mathcal{P}_{\Uin, \mathrm{id}}}[1-2x] }  \text{ .}
\end{equation*}
Hence, we can obtain a $\tau$-close approximation of the expectation value if the white noise probability $\epsilon$ is upper-bounded as shown above. Note that we chose a high-level noise model where the white noise are i.i.d. bit flips. We leave the investigation of other quantum noise models to obtain bounds for $\tau$ to future work. 

\begin{figure*}[]
\centering
\includegraphics[width=0.83\textwidth]{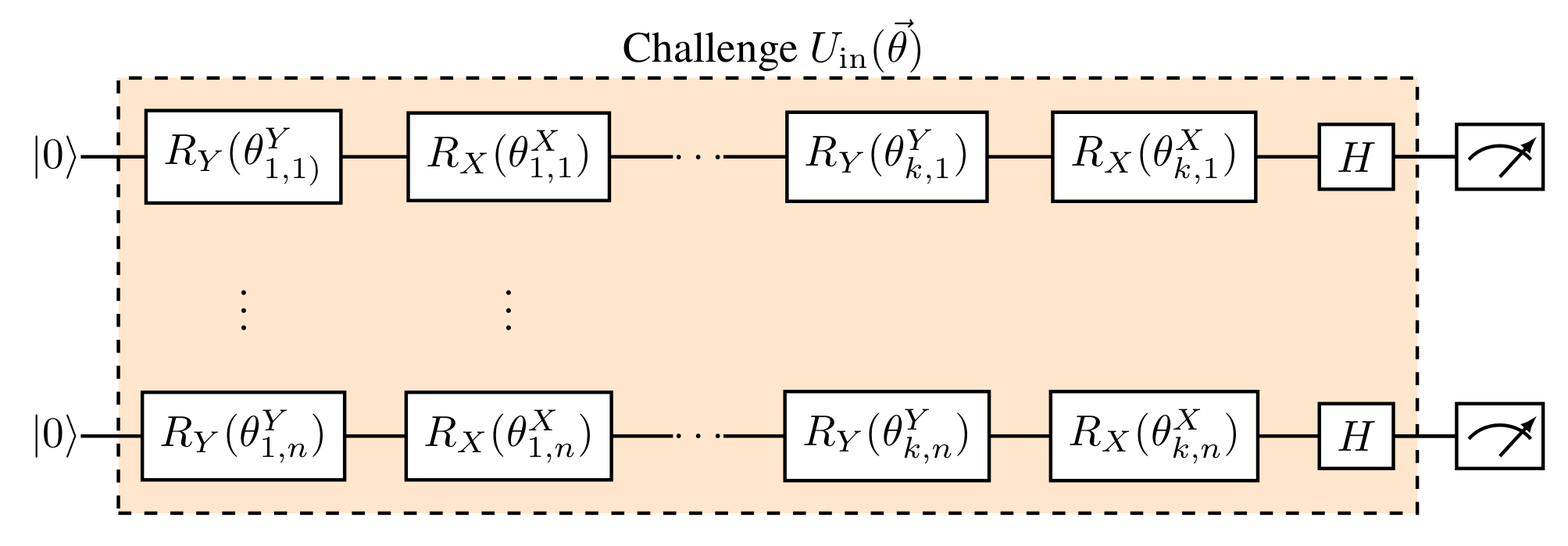}
\caption{Circuit of the parameterized challenge unitary family as a natural extension to Hadamard CR-QPUF. The circuit consists of $k$ layers of parameterized $R_{Y}$ and $R_{X}$ rotations and a Hadamard gate that are performed on $n$ qubits.}
\label{fig:h-qupuf_circuit_multi}
\end{figure*}


\cite{phalak_quantum_2021} propose to obtain multiple responses for every challenge and to map every component in each response $\Vec{R}_{\mathrm{out}}$ to a $5$-bit string resulting in the $(5\cdot n)$-bit string $S_\mathrm{out}$, which make up the CRP database.
The protocol then accepts the respective string $S'_\mathrm{out}$ of a response if the average Hamming distance between $S'_\mathrm{out}$ and all respective strings to the same challenge in the CRP database is within the variance among the strings in the CRP database.
The authors chose the $5$-bit representation as a favorable trade-off between uniqueness of the signature and robustness to white noise.

\subsection{Extension to multiple rotations}
\label{sec:extensions-intro}

A natural extension of the Hadamard CR-QPUF as proposed by \cite{phalak_quantum_2021} is to increase the number of rotations applied to the qubits to \textit{rotation chains}.
For example, consider challenges of the form
\begin{equation*}
    U_{\mathrm{in}}(\Vec{\theta_{1}}, ..., \Vec{\theta_{t}}) = \bigotimes_{j=1}^{n} H R_{Y}(\theta_{1,j}) R_{X}(\theta_{2,j})...R_{Y}(\theta_{t,j})
\end{equation*}
that consists of a chain of alternating $R_{Y}$ and $R_{X}$ rotations (figure \ref{fig:h-qupuf_circuit_multi}). In our modelling attack on the Hadamard CR-QPUF we also consider such natural extensions and show that the resulting responses can be learned using a simplified model. In the simplified model, the responses of CR-QPUFs based on arbitrary rotation chains consisting of $R_{Y}$ and $R_{X}$ rotations are learned. Therefore, a model using only two rotations, i.e. one $R_Y$ and one $R_X$ rotation is learned.

\section{Attacker Capabilities}

In this section we distil the capabilities an attacker must have to perform the modelling attack. One application of CR-QPUFs is to protect quantum cloud customers from fraudulent resource allocation on the provider side.
One such example is when a customer is assigned by a malicious actor, to cheaper and lower-grade quantum computing hardware that what they had originally agreed upon. In particular, this includes that the cloud provider is malicious and aims at assigning the customer to lower-grade hardware.
In the attacker model we consider, the malicious actor knows the challenge space $\mathcal{U}$ and is able to run chosen challenges $\Uin \in \mathcal{U}$ on the quantum computer that she aims to model. Additionally she is able to detect when a user
tries to run a Hadamard CR-QPUF challenge and extract the challenge rotations $\Vec{\theta}$.
This is well in the threat model of a malicious quantum cloud provider. It is usually assumed that an attacker of PUF authentication schemes knows the challenge space and tries to predict the response to an unknown challenge.
In our case, the attacker will predict the correct response according to $\Vec{\theta}$ using a model that was learned in a learning phase.\\
As we will show in the following section, such a malicious actor that can reroute user circuits to lower-grade, remains completely undetected by the Hadamard CR-QPUF secure provisioning protocol.

\section{The Modelling Attack}

The attack on the Hadamard CR-QPUF secure provisioning scheme is carried out
in two phases, the learning phase and the attack phase. One important step in the attack is based on the observation that the
Hadamard CR-QPUF does not entangle the single qubits and thus creates the opportunity to learn the characteristics of the qubits individually. The second essential observation is the fact that the challenge unitaries are fixed in their circuit structure, hence the challenge space can be reduced to the parametric rotations $\Vec{\theta}$. In the following we describe the learning phase, where the design shortcomings of the Hadamard CR-QPUF are exploited to learn a model of the CR-QPUF, and the attack phase, where the actual attack using the learned models is carried out.

\subsection*{Learning Phase}

In the learning phase, the attacker Eve gathers $L$ CRPs $\{ ( \Uin(\Vec{\theta_l}),  \Vec{R}_\mathrm{out}) : \Vec{\theta_l} \in [0,2\pi)^n \text{, } l=1,...,L\}$, where the chosen rotations $\Vec{\theta_l}$ cover the space $[0,2\pi)^n$ equidistantly, i.e. in a grid. Here the component $\theta_{l,i} = \theta_{l,j}$ for one $\Vec{\theta_l}$ for $i,j \in \{1,...,n\}$, which means that all $n$ qubits are rotated by the same angle.\\
Having obtained the responses for the chosen rotations, she then fits a bounded degree polynomial $f^{(j)}$ to the obtained
data points $\{ ( \theta_{l,j}, {r_l}_\mathrm{out}) : l=1,...,L \}$ for each qubit $j$. The degree of the polynomials $f^{(j)}$ are upper bounded by $1/\epsilon$, where $\epsilon$ is the error probability of the qubit \citep{benjamini1999noise}. We estimate the error probability of the qubits to be $0.1$ and thus choose a degree of 10 for the polynomials $f^{(j)}(\theta_j)$.\\
The fitting of the polynomials is done using a textbook least square error regression using the \texttt{scikit-learn} library \citep{scikit-learn}.
Using this regression, Eve learns a model function $f^{(j)}(\theta_j)$
for each qubit $j\in\{1,...,n\}$ that predicts the $j^{th}$-component of $\Vec{R}_\mathrm{out}$.
Given unknown challenge rotations $\Vec{\theta}$, the learned model functions will then be used in the attack phase to calculate the predicted response $\Vec{R}'_\mathrm{out}$, where:
\begin{align*}
    \Vec{R}'_\mathrm{out} &= f(\Vec{\theta}) =  
        \begin{pmatrix}
           f^{(1)}(\theta_1) \\           
           \vdots \\
           f^{(n)}(\theta_n)
    \end{pmatrix}.
  \end{align*}

\subsection*{Attack Phase}

\begin{figure}[h]
\begin{centering}
\includegraphics[width=1\columnwidth]{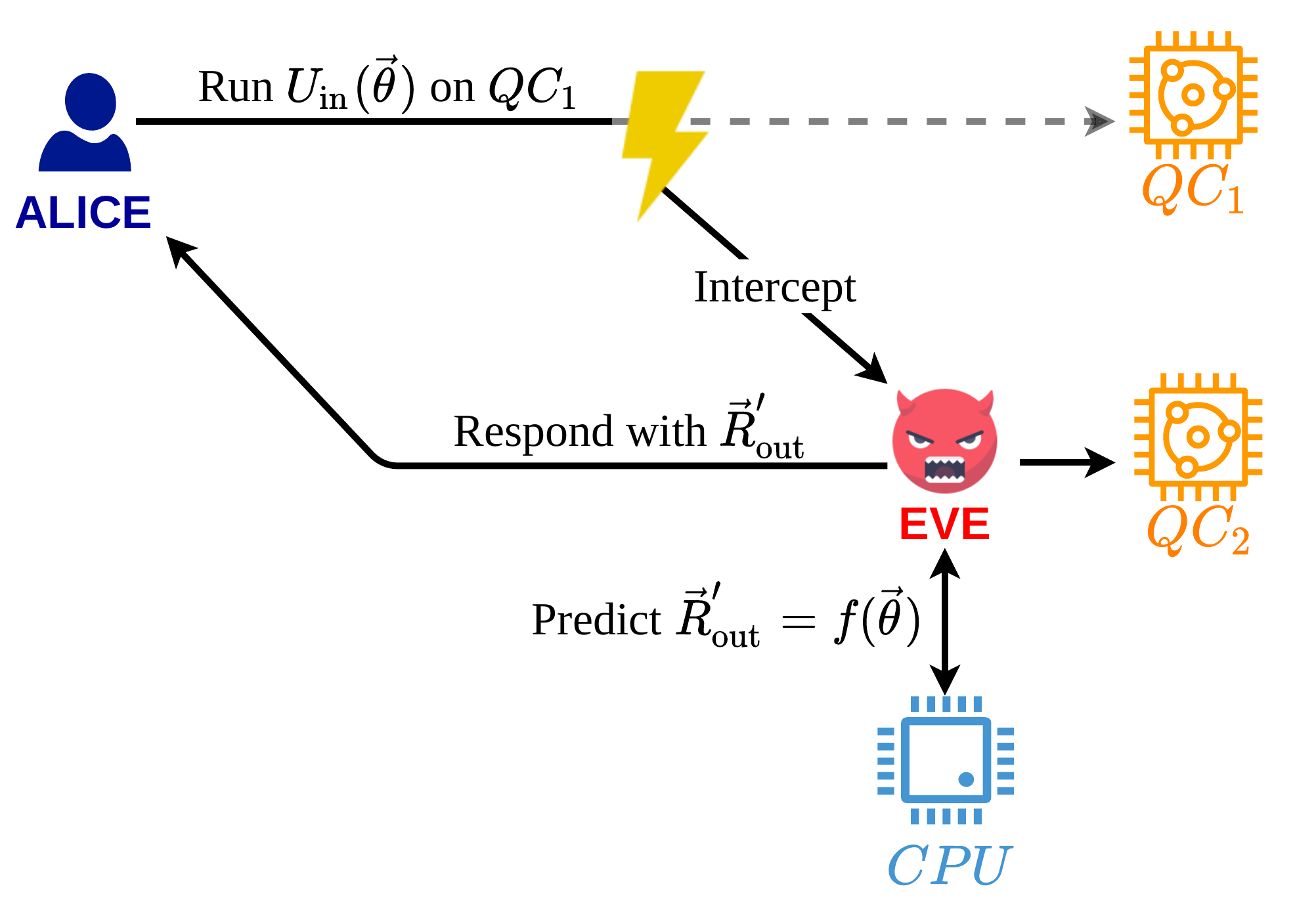}
\par\end{centering}
\caption{The malicious actor Eve on the QC cloud provider side runs all circuits of Alice on lower-grade hardware $QC_{2}$ and intercepts queries to the Hadamard CR-QPUF. Eve
will use the model function $f(\Vec{\theta}) = \Vec{R}'_\mathrm{out}$ to predict $\Vec{R}_{\mathrm{out}}$ and send the prediction back to Alice.}
\label{fig:attack_phase}
\end{figure}

In the attack phase, Eve intercepts the requests where Alice
wants to run a circuit on $QC_{1}$ and runs them on a lower-grade machine $QC_{2}$.
Whenever Alice runs a Hadamard CR-QPUF challenge $\Uint$ to detect mal-provisioning,
Eve responds with the prediction $R'_\mathrm{out}$ using the regression
models $f^{(j)}(\theta_j)$ that have been learned in the learning phase. The
attack layout is illustrated in figure \ref{fig:attack_phase}.
If Alice accepts the response $\Vec{R}'_\mathrm{out}$, Eve successfully
tricked Alice into falsely authenticating $QC_{1}$, while in fact,
all circuits of Alice have been run on $QC_{2}$. In what follows, we will show how this attack performs using a real-world commercial quantum cloud computer provided by IBM.

\section{Results}
\label{sec:results}

We carried out the learning and attack phase in a real-world scenario using the 27-qubit IBM quantum cloud computer $\texttt{{ibmq\_mumbai}}$.
Firstly, we learned the model function $f^{(j)}$ for each qubit $j=1,...,27$.
Therefore we ran the Hadamard CR-QPUF challenge unitary $U_{\mathrm{in}}(\Vec{\theta_l})$ for $L=30$ different rotations $\Vec{\theta_l}$ on $\texttt{{ibmq\_mumbai}}$ and obtained the SQ responses $\Vec{R}_{\mathrm{out},l}$ by averaging over $\abs{\hat{r}_{\mathrm{out},l}} = 2000$ shots, where $l = {1,...,L}$.
The rotation angles $\Vec{\theta_{l}}$ were chosen such that the challenge input space $[0,2\pi)$ is split equally in $L$ angles.
\begin{figure}[h]
\begin{centering}
\includegraphics[width=1\columnwidth]{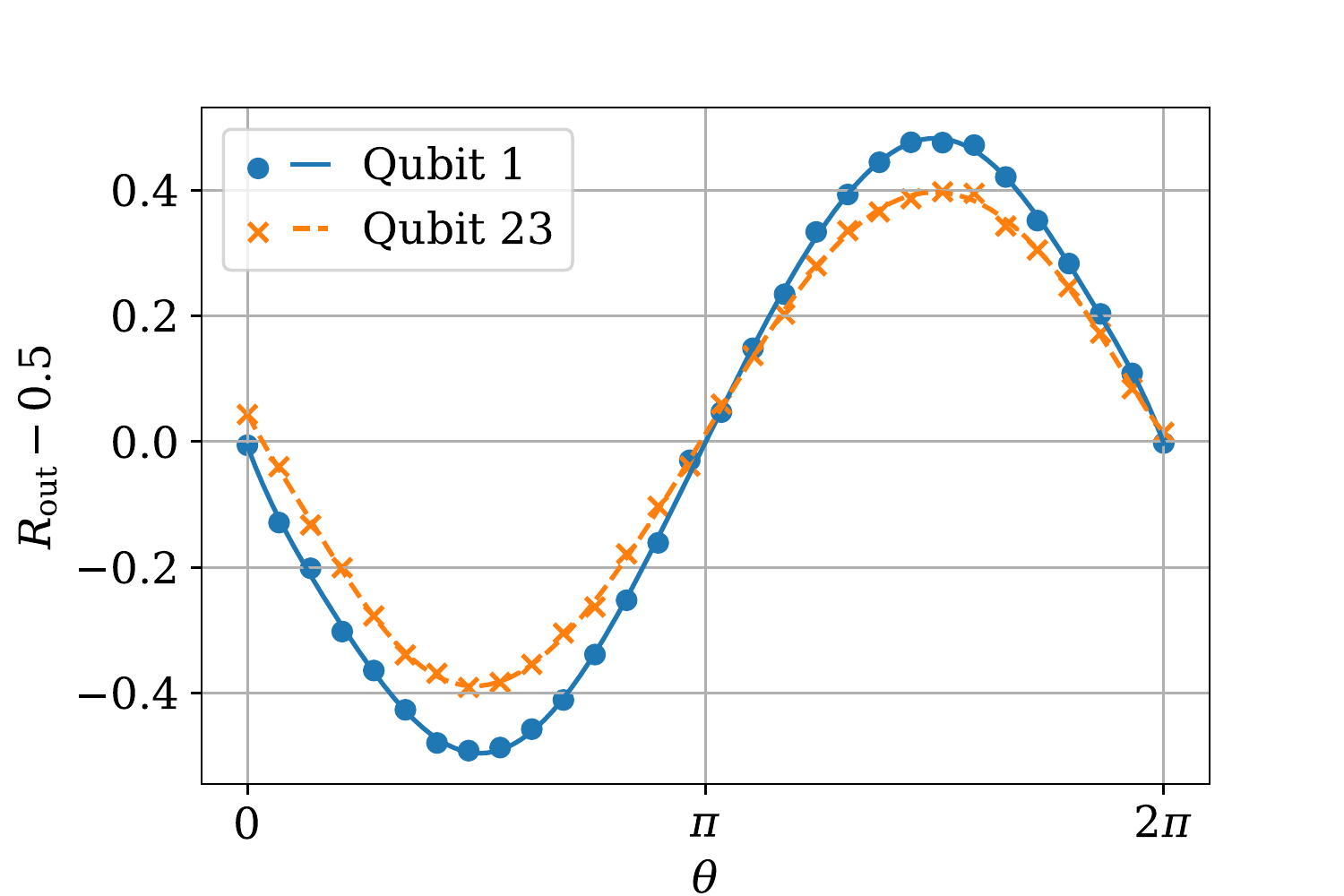}
\par\end{centering}
\caption{Measured responses along the $30$ rotation angles between
$0$ and $2\pi$ for two different qubits of the $\texttt{{ibmq\_mumbai}}$ commercial
quantum computer. The interpolated lines are the fitted prediction models for the respective qubits. The two qubits exhibit different characteristics depending on $\theta$.
The dependence of the responses on the challenges can be accurately modelled using the low-degree polynomials.}
\label{fig:biases_two_qubits}
\end{figure}\\
Figure \ref{fig:biases_two_qubits} shows the $L=30$ measured responses
for two different qubits and the respective regression models
$f^{(j)}(\theta_j)$ that are learnt from the data samples $\{ ( \theta_{l,j}, \Vec{R}_{\mathrm{out},l,j}) : l = 1,...,30 \}$
for qubits $j=1,23$. The plot depicts how the responses for individual qubits are degenerated from a perfect sinus function, which is the optimal analytical solution for the responses. This degeneration is due to imperfections of the quantum computer. 

\begin{figure}
\begin{centering}
\includegraphics[width=1\columnwidth]{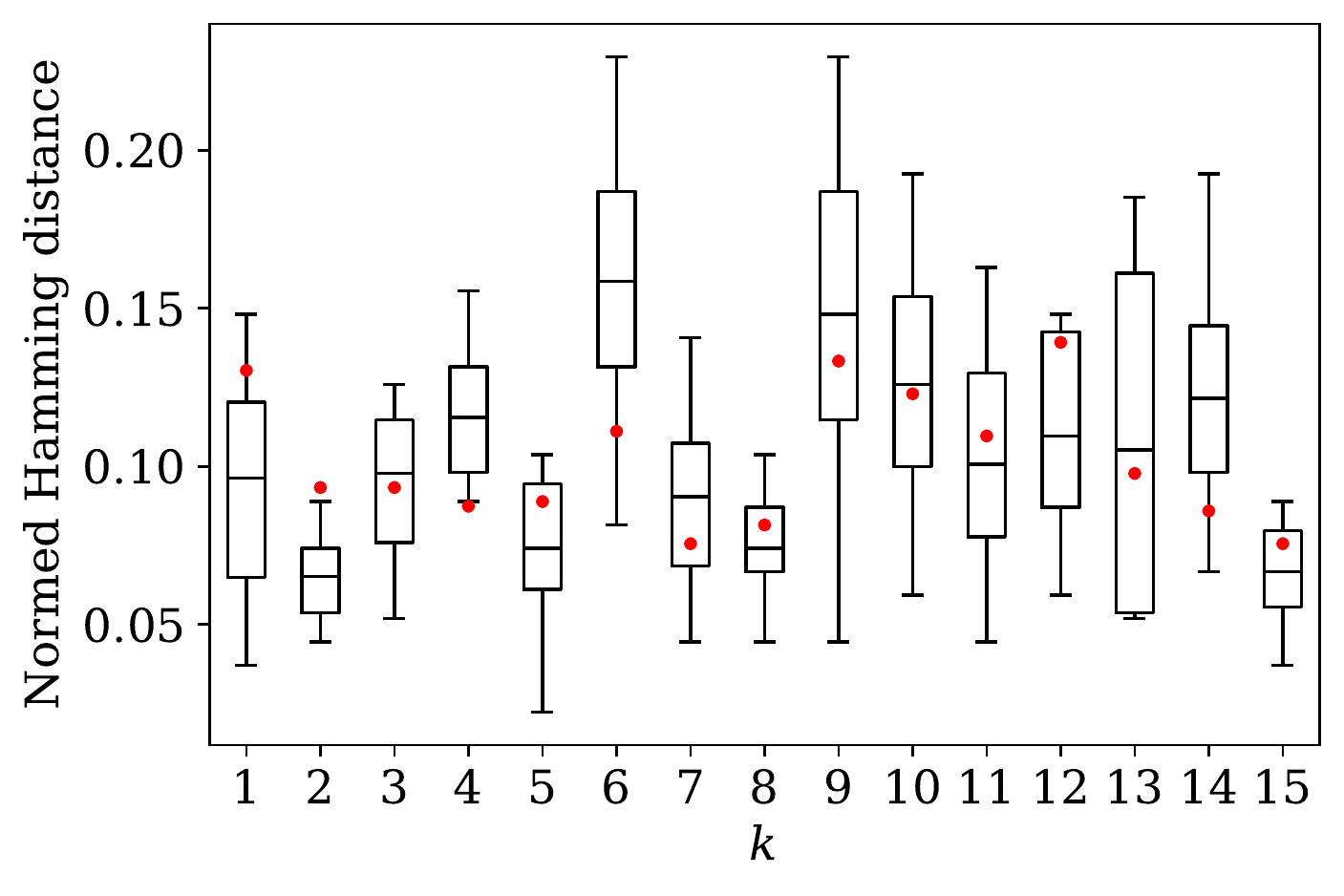}
\par\end{centering}
\caption{Box plot showing the variance of the Hamming distances of responses in the holdout CRP database and their mean. The inner boxes resemble the middle two quartiles of the data. The red dots show the average HD of the $k=15$ predicted responses (obtained through the model functions $\{f(\Vec{\theta}_k):k=1,...,15\}$) to the respective responses in the CRP database.
Almost every predicted response will get accepted by the authentication protocol since their average HDs to the respective responses in the CRP database are within the variance intervals.}
\label{fig:HD_AFA}
\end{figure}

In the attack phase, we randomly sampled $15$ challenges $\{ \Uin(\Vec{\theta_k}) : k=1,...,15\}$ (each $5$ times) to build a holdout CRP database. We then calculated the respective predicted responses $\{ f(\Vec{\theta_k})) : k=1,...,15 \}$ using the learned regression models.
After mapping the predicted responses to $25$ bits, as suggested by \cite{phalak_quantum_2021}, we calculated their average Hamming distance (HD) to the holdout CRP database.
Per the protocol, if the average HD is within the variance interval of the HDs of the respective responses in the database, the predicted response gets accepted.\\
In figure \ref{fig:HD_AFA}, we plot the intra HD of the (holdout) CRP database and the respective predicted responses which we obtained in the attack. As we can see, almost all of the predicted responses get accepted.
Hence, we were able to model the Hadamard CR-QPUF behaviour and successfully predicted
responses to unknown challenges. This showcases the
predictability and thus the insecurity of the Hadamard CR-QPUF.
An attacker on the QC cloud provider side, that is able to detect
when the Hadamard CR-QPUF challenge is requested to run, can successfully trick
the authentication scheme. It will enable the attacker to reroute
client circuits to lower-grade hardware without detection.\\

\subsection{Extension to multiple rotations}

\begin{figure}
    \centering
    \includegraphics[trim=0 0 0 0, scale=.4]{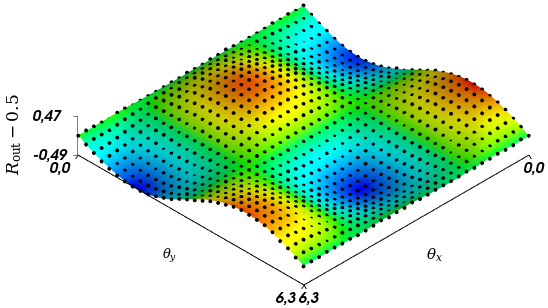}
    \caption{Training samples (black dots) and the landscape of $f^{(j)}(\theta_X, \theta_Y)$ (surface) that has been learned for one qubit.
    The landscape exposes a mixture of a cosine and a sine function in the $\theta_X$ and $\theta_Y$ directions, respectively.
    The learned model will be used to imitate the characteristics of the qubit for multiple $X$ and $Y$ rotations.}
    \label{fig:2d-model}
\end{figure}

\begin{figure*}
    \centering
    \includegraphics[width=1\columnwidth]{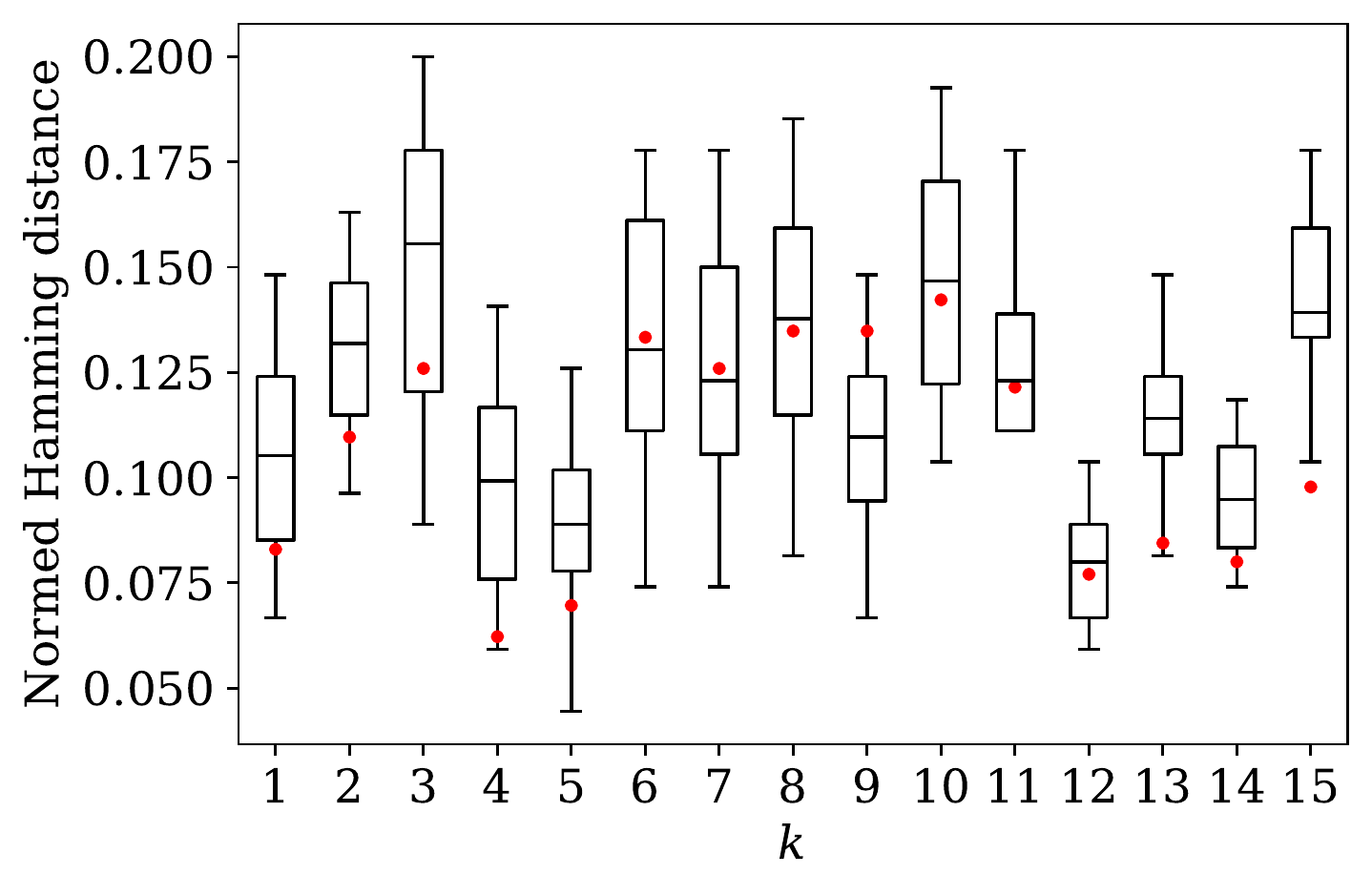}
    \includegraphics[width=1\columnwidth]{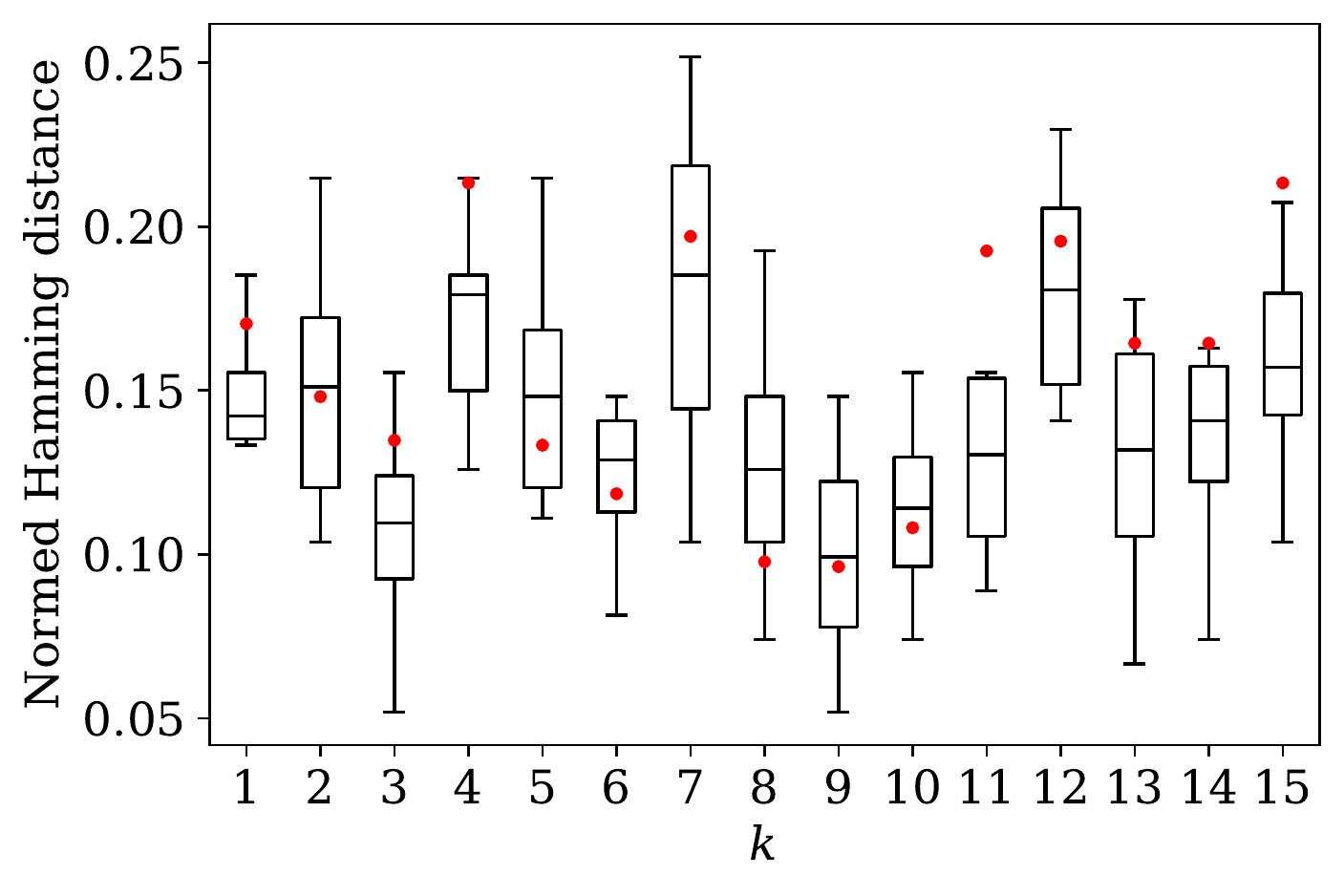}
    \caption{
    Box plot showing the variance of the Hamming distances of responses in the holdout CRP database and their mean. The inner boxes resemble the middle two quartiles of the data. The red dots show the average HD of the $k=15$ predicted responses (obtained through the model functions $\{f(\Vec{\theta}_k):k=1,...,15\}$) to the respective responses in the CRP database.
    Almost every predicted response will get accepted by the authentication protocol since their average HDs to the respective responses in the CRP database are within the variance intervals.
    Challenges of the form $H R_{Y}R_{Y}R_{X}R_{X}$ (left) and $H R_{Y}R_{X}R_{Y}R_{X}R_{Y}R_{X}R_{Y}R_{X}$ (right) were employed.
    }
    \label{fig:2d-results}
\end{figure*}

In this subsection we present the results of our attack on natural extensions of the Hadamard CR-QPUF as described in section \ref{sec:extensions-intro}. Here, the challenge unitaries do not consist of only one, but of multiple $X$ and $Y$ rotations per qubit, forming a chain of rotations.
The idea of the attack is again to learn a simplified polynomial that only depends on two rotations $\theta_X$ and $\theta_Y$ for each qubit.\\
In the learning phase, we learn a polynomial $f^{(j)}(\theta_{X,j}, \theta_{Y,j})$ of degree $10$ for each qubit $j$ that depends on only $X$ and $Y$ rotations.
To learn the models, we obtained training samples by running the challenge
\begin{equation*}
    U_{\mathrm{in}}(\Vec{\theta}_{X}, \Vec{\theta}_{Y}) = \bigotimes_{j=1}^{n} H R_{Y}(\theta_{Y,j}) R_{X}(\theta_{X,j}) 
\end{equation*}
using rotation angles $\Vec{\theta}_{X}$ and $\Vec{\theta}_{Y}$ that equidistantly cover the space $[0,2\pi) \times [0,2\pi)$ in a discretized $30 \times 30$ grid.
Figure \ref{fig:2d-model} shows the training samples obtained from \texttt{ibmq\_mumbai} and the landscape of one of the trained models $f^{(j)}(\theta_X, \theta_Y)$.
\\
In the attack phase, we aim at predicting responses to unknown challenges comprising of chains of rotations.
In order to predict the response to a challenge $U_{\mathrm{in}}(\Vec{\sigma}_{\rho_1}, ..., \Vec{\sigma}_{\rho_t})$, where $\rho_1,...,\rho_t \in \{X,Y\}$ are the rotation directions and the $\Vec{\sigma}$'s are the rotation angles for all qubits, we sum over the $\Vec{\sigma}_X$'s and $\Vec{\sigma}_Y$'s in the challenge and pass them into the learned qubit models $f^{(j)}(\theta_{X,j}, \theta_{Y,j})$. More precisely:
\begin{align*}
    \Vec{\theta}_{X} &= \sum_{i=1, \rho_i = X}^{t} {\Vec{\sigma}_{\rho_i}} \mod 2\pi\\
    \Vec{\theta}_{Y} &= \sum_{i=1, \rho_i = Y}^{t} {\Vec{\sigma}_{\rho_i}} \mod 2\pi
\end{align*}
\begin{equation*}
    \Vec{R}'_\mathrm{out} = f(\Vec{\theta}_{X}, \Vec{\theta}_{Y}) =  
        \begin{pmatrix}
           f^{(1)}(\theta_{X,1}, \theta_{Y,1}) \\           
           \vdots \\
           f^{(n)}(\theta_{X,n}, \theta_{Y,n})
        \end{pmatrix}
\end{equation*}\\
Analogously to section \ref{sec:results}, we carried out the attack on a real-world quantum computer.
In figure \ref{fig:2d-results} we present the results of this modelling attack on two different kinds of extensions to the Hadamard CR-QPUF. The attack was executed again on the $27$-qubit commercial quantum computer \texttt{ibmq\_mumbai}. As one can see, again almost all of the predicted responses will get accepted by the protocol, showcasing the insecurity of these extensions of the Hadamard CR-QPUF in the SQ model.

\section{Discussion}

In this work, we have examined the class of Classical Readout Quantum PUFs. When designing CR-QPUF schemes and in the presence of noisy devices, it is natural to construct robust responses by using the expectation value of a random variable, which is then approximated using multiple noisy samples. This formalism is known as the statistical query model \citep{kearns98}. Recently, \cite{learning-local-quant-dist} found that output distributions of local quantum circuits cannot be learned efficiently when the learner is restricted to the SQ model (both for classical and quantum learners).
Other results \citep{gollakota2021hardness} show that learning stabilizer states in the presence of noise and in the SQ model is intractable.

In the context of noisy quantum devices, this situation poses a very interesting perspective:
since NISQ devices are inherently noisy, can one leverage the restriction to the SQ model to derive provable security against learning attacks?
For instance, can one construct a CR-QPUF, such that one can show that learning the CR-QPUF in the SQ model implies learning an unknown stabilizer state in the SQ model or an unknown local quantum circuit in the SQ model?
We believe this poses a very interesting research question. If the answer would be yes, then provable security guarantees against learning attacks could be derived for the CR-QPUF on imperfect noisy devices.

Nevertheless, there are substantial uncertainties that need to be addressed.
Firstly, since the secret in CR-QPUFs are the device imperfections, can the device specific imperfections be modelled and learned?
This corresponds to learning $\qpufid$ which would mean that the underlying security parameter $\lambda$ can be reconstructed and render any QPUF based on $\lambda$ useless. As of now, to the best of our knowledge, there are no results that show that the device imperfections cannot be accurately modelled. We believe that in order to resolve this question, results in the field of quantum process tomography \citep{Altepeter2003, Mohseni2008} might help us to better understand whether device imperfections can be generally modelled or whether given $\Uin \in \mathcal{U}$ the device specific born distribution $\mathcal{P}_{\Uint, \mathrm{id}}$ cannot be learned in the \emph{average} case.
Secondly, since the challenge unitary $\Uin$ is known to the attacker, do the device imperfections alter the output distribution such that knowing $\Uin$ does not constitute a significant attacker advantage in learning the output distribution? Can the influence of the imperfections be predicted knowing the challenge unitary?
These questions need to be clarified, to be confident about the security against learning attacks in the SQ model.

Another important aspect to keep in mind is that an attacker with direct access to the QC cloud provider machine (in contrast to our above pure communication interception based attacker) is generally able to act outside of the SQ model. 
Even though there are learning problems where learning with noise is a hard task (even for quantum computers \citep{gollakota2021hardness}), the extended attacker is clearly not limited to the SQ model. 
In this less restricted learning scenario direct usage of the noisy samples $\rhat$ is clearly possible. 

However, studies have also shown that for the only known separation result between the SQ model and the PAC model (for the learning parity with noise problem), there is only a tiny advantage for using noisy samples over the SQ model \citep{blum2003noise}.
This gives strong reason to believe that for hard learning problems in the presence of noise, acting outside of the SQ model does not yield a significant advantage.

If one wants to take a very pessimistic stance to quantum computing, fundamentally, according to \cite{kalai2020argument} and others, due to the excessive noise in NISQ devices, the complexity of these devices does not exceed the class of low-degree polynomials at all. 
This means that NISQ machines are computationally not stronger than classical computers and classical computers are in theory able to simulate calculations on NISQ devices.
This again would of course be detrimental to CR-QPUFs, since any CR-QPUF would be efficiently learnable using low-degree polynomials. While this argument is heavily debated in the quantum computing community at this time, designing CR-QPUFs and testing their security thereof provides an excellent playground to check "the argument against quantum computers" and challenge the NISQ expressivity.
CR-QPUF proposals could  be checked against classical and quantum learning attacks, providing evidence to the question whether NISQ computations form a low-complexity class of algorithms whose output can be learned using low-degree polynomials.

As we have found in the insecurity of the Hadamard CR-QPUF, the responses can be learned using low-degree polynomials, which shows us that the class of challenge unitaries needs to be designed very carefully. Since the Hadamard CR-QPUF does not use entanglement, the QPUF security reduces to that of one single qubit and an attacker only needs to learn the characteristics of one qubit at a time.
We therefore argue that entanglement is strictly required in CR-QPUF challenge choices.
Additionally, when designing a CR-QPUF, one can leverage that the class of challenge unitaries is not fixed in their circuit structure. In the Hadamard CR-QPUF proposal, this property is not leveraged and the challenge circuit is structurally fixed, which reduces the challenge space to the single qubit rotation angles $\Vec{\theta}$.
Finally we want to mention that while the idea of using imperfections in quantum computers to create an unforgeable fingerprint of the devices could potentially be used in QPUF schemes, opposing incentives of quantum computer manufacturers, who want to eliminate imperfections, and QPUF users, who require imperfections to identify devices, might hinder their application to industrial cloud providers.

\section{Future Work}
Going forward, analyzing and categorizing the device imperfections and their influence on degenerating the Born distribution output would serve a better understanding of CR-QPUFs. 
In particular, can device imperfections be leveraged such that the mutation of the output distribution of local quantum circuits is not learnable in the SQ model? 
And equally importantly, can the device imperfections be leveraged such that their influence remains secret and unpredictable for future challenge unitaries? We believe that proposing justified CR-QPUF designs and testing their security, using tools developed for quantum process tomography and machine learning, would help to gather evidence of the prospects and feasibility of QPUFs. 
This poses as a excellent playground to gain further insight to CR-QPUFs based on quantum device imperfections.

\backmatter

\bmhead{Acknowledgments}
The authors partially acknowledge funding by the Einstein Research Unit "Perspectives of a quantum digital transformation: Near-term quantum computational devices and quantum processors" of the Berlin University Alliance. A.\ Pappa acknowledges support from the German Research Foundation (DFG, Emmy Noether grant No. 418294583). We thank Prof. Elham Kashefi and her research group at the University of Edinburgh for the helpful discussions.

\section*{Declarations}

\bmhead{Funding}
Not applicable

\bmhead{Conflict of interest}
The authors declare no conflict of interests.

\bmhead{Ethics approval}
Not applicable

\bmhead{Consent to participate}
Not applicable

\bmhead{Consent for publication}
Not applicable

\bmhead{Availability of data and materials}
Not applicable

\bmhead{Code availability}
The code used in this work is available at \url{https://github.com/n1kn4x/H-CRQPUF_Attack}.

\bmhead{Authors' contributions}
Not applicable


\bibliography{qupuf_attack_springer.bbl}



\end{document}